\documentclass[11pt]{article}
\pdfoutput=1
\usepackage[margin=0.8in]{geometry}
\usepackage{amsmath,amssymb,graphicx}
\usepackage{hyperref}
\usepackage{slashed}
\usepackage{verbatim}

\usepackage[T1]{fontenc}
\usepackage[utf8]{inputenc}

\usepackage{titlesec}
\renewcommand{\thesection}{\Roman{section}}
\titleformat{\section}[block]{\bfseries\centering}{\thesection.}{1em}{}
\renewcommand{\thesubsection}{\arabic{subsection}}
\titleformat{\subsection}[block]{\bfseries\centering}{\thesection.\thesubsection}{1em}{}

\usepackage[font=small,labelfont=bf]{caption}
\usepackage{cite}

\newcommand{\be}{\begin{equation}}
\newcommand{\ee}{\end{equation}}

\usepackage{xcolor}
\usepackage{bm}

\title{\bf  Snowmass White Paper: \\
Implications of Quantum Gravity\\ 
for Particle Physics}

\newcommand\snowmass{\begin{center}\rule[-0.2in]{\hsize}{0.01in}\\\rule{\hsize}{0.01in}\\
\vskip 0.1in Submitted to the  Proceedings of the US Community Study\\ 
on the Future of Particle Physics (Snowmass 2021)\\ 
\rule{\hsize}{0.01in}\\\rule[+0.2in]{\hsize}{0.01in} \end{center}}

\begin{document}

\maketitle
\thispagestyle{empty}

\begin{center}
\author{Patrick Draper$^{(a) \, *}$, Isabel Garcia Garcia$^{(b) \, \dagger}$, and Matthew Reece$^{(c) \, \ddagger}$}
\end{center}

\begin{center}
$^{(a)}$ \emph{Department of Physics and the Illinois Center for the Advanced Study of the Universe,\\ University of Illinois, Urbana, IL 61801} \\
$^{(b)}$ \emph{Kavli Institute for Theoretical Physics, University of California, Santa Barbara, CA 93106} \\
$^{(c)}$ \emph{Department of Physics, Harvard University, Cambridge, MA 02138}
\end{center}

\bigskip

\begin{center}
$^*$pdraper@illinois.edu,
$^\dagger$isabel@kitp.ucsb.edu,
$^\ddagger$mreece@g.harvard.edu
\end{center}

\bigskip

\begin{abstract}
	Quantum gravity places important consistency conditions on low-energy effective field theory, such as the absence of global symmetries. These may have important consequences in the search for particle physics beyond the Standard Model. We review some of these conditions and their phenomenological implications for the strong CP problem, the weak scale, new gauge interactions, and cosmology. We also offer some general comments on how these ideas can guide model building.
\end{abstract}

\snowmass

\newpage
\pagenumbering{arabic}

\section{Introduction}
\label{sec:intro}

The framework of Effective Field Theory (EFT) has played a crucial role in understanding the dynamics of quantum field theories. Yet we know that the rules of EFT must bend, if not break, in a theory that consistently incorporates gravity. From the absence of global symmetries to the sub-extensive entropy of black holes, there is unambiguous evidence that the rules of EFT-consistency as currently understood are not enough to describe even the low energy regime of theories that admit a gravitational UV-completion.

Can one guarantee that gravitational considerations beyond EFT will play a role in resolving some of the long-standing problems in particle physics, such as, say, the electroweak hierarchy, the cosmological constant, or the strong CP problems? No. But it is an interesting possibility that they do, and the goal of this whitepaper is to review arguments that exploring this possibility is a promising direction in particle theory, with the potential to shed new light on -- at least -- the properties of quantum field theories coupled to gravity and -- at best -- on some of the structural aspects of our Universe that remain mysterious.

In  section~\ref{sec:gravityvsEFT}, we provide a general  overview of how quantum gravity constraints EFTs. In section~\ref{sec:particlephysics}, we focus on how these general principles relate to specific applications in particle physics. Considerations related to dark energy and the cosmological constant are discussed in section~\ref{sec:darkenergy}. We conclude briefly in section~\ref{sec:conclusions}.

\section{Gravity vs Effective Field Theory}
\label{sec:gravityvsEFT}

\subsection{No global symmetries}

Theories of quantum gravity, in contrast to quantum field theories, are expected to lack any form of global symmetry. Heuristically, this has long been believed to be the case because black holes appear to destroy global charges -- an argument that has been expanded and refined over time~\cite{Hawking:1975vcx, Zeldovich:1976vq, Giddings:1988cx, Coleman:1989zu, Abbott:1989jw, Kallosh:1995hi, Banks:2010zn, Harlow:2018tng, Harlow:2020bee, Chen:2020ojn, Hsin:2020mfa}. Conceptually, this observation is at odds with the idea of ``technical naturalness'' introduced by 't Hooft \cite{tHooft:1979rat}: resting on the observation that the symmetry structure of a theory at the classical level is respected by quantum corrections in a quantum field theory, it postulates that a dimensionless parameter $\alpha$ is justified to be $\ll 1$ provided the limit $\alpha \rightarrow 0$ restores a symmetry of the theory.
Within the framework of EFT, 't Hooft's principle has played a major role in model building in physics beyond the Standard Model, often used as the ultimate consideration to decide whether extensions of the Standard Model are well-motivated from a theoretical standpoint.

It is clear that a more sophisticated version of 't Hooft naturalness is necessary in the presence of gravity. Achieving this goal necessitates quantitative progress: exactly \emph{how much} does gravity violate global symmetries? For example, baryon number is an accidental symmetry of the Standard Model, with no renormalizable operators that break it. However, there exist dimension six operators that violate baryon number, such as $qqql$ and $uude$. A general expectation is that these appear with coefficients at least of order $1/M_\mathrm{Pl}^2$, mimicking the effect of integrating out degrees of freedom with symmetry-violating interactions present at the gravitational cutoff. One might wonder if this paradigm is universal: should all global symmetries be broken by Planck-suppressed higher dimension operators, so that low-energy symmetry violating effects are suppressed by {\em powers} of $E/M_\mathrm{Pl}$ or $m/M_\mathrm{Pl}$ (with $m$ an IR mass scale)?

The answer to this question is {\em no}, and such reasoning can lead one astray in two different directions. On the one hand, it is possible for symmetry violation in the IR to be {\em exponentially small}, a point to which we will return shortly. On the other, it is sometimes assumed in model-building that a global symmetry holds at the renormalizable level, broken only by Planck-suppressed operators. In other words, it is assumed that one can {\em impose} a global symmetry, as one would in quantum field theory, and then add effects ``of gravitational size'' to capture the symmetry breaking. But this is not expected to ever be correct: if a symmetry is explicitly broken, there is no  reason for relevant or marginal symmetry violating terms to be absent. They should be expected to appear with $O(1)$ coefficients, {\em unless} such contributions can be forbidden by some underlying (possibly discrete, possibly spontaneously broken) gauge symmetry, or by some other fundamental physics such as locality in extra dimensions. In other words, {\em approximate global symmetries exist only as artifacts enforced by other principles in the UV}. The underlying principles might be hidden, from the IR point of view, just as one working at the scale of pion physics finds isospin to be an approximate symmetry not explained by any underlying structure. Within the electroweak theory, we make progress in explaining the origin of approximate isospin in terms of the fact that  the up and down quark masses are forbidden by electroweak symmetry, and appear only below the scale of Higgsing. However, it also requires that the up and down quark Yukawas are small, a feature that we must go beyond the Standard Model to explain, e.g., in terms of new  gauged flavor symmetries or extra-dimensional locality. At the high energy scale where gravity becomes strongly coupled, we expect that there are no approximate symmetries whatsoever; every approximate IR symmetry is expected to be badly broken in the UV~\cite{Nomura:2019qps,Cordova:2022rer}.

Attempts to find a universal upper bound on the size of global symmetry violation in a theory of quantum gravity are as old as the conjecture itself \cite{Coleman:1989zu,Kallosh:1995hi}, and have received renewed attention in recent times \cite{Fichet:2019ugl, Daus:2020vtf}. The current consensus can be summarized by saying that the size of symmetry-violating effects in the infrared is expected to be at least of order $\exp(-8\pi^2 M_\mathrm{Pl}^2/\Lambda^2)$, where $\Lambda$ is a UV cutoff. This could correspond to the quantum gravity cutoff, but more general arguments suggests it could also signal the onset of a less dramatic UV-completion \cite{Fichet:2019ugl, Daus:2020vtf}. A very heuristic argument for this is that we expect black holes to violate global symmetries, so one might suspect the minimum violation is of order $\exp(-S_\mathrm{BH;min}) = \exp(-8\pi^2 M_\mathrm{Pl}^2 R_\mathrm{BH;min}^2)$ where $R_\mathrm{BH;min}$ is the radius of the smallest semiclassical black hole.

The expectation that global symmetry violation may be exponentially tiny is consistent with ingredients in string theory constructions that allow for concrete realizations of approximate global symmetries. For example, axion-like fields can have an approximate continuous global shift symmetry. It is explicitly broken by an axion coupling to gauge fields, but this manifests in an exponentially small, non-perturbative potential of order $\exp(-8 \pi^2 / g^2)$. Similarly, the shift symmetry of axions arising as zero modes of higher-dimensional gauge fields upon compactification may be broken by Chern-Simons couplings or couplings to matter or branes charged under the higher-dimensional gauge field, but these again manifest in potentials that are exponentially suppressed when the compactification volume is large in units of the inverse tension of the charged objects. Another example arises in intersecting D-brane models in type IIA string theory, where Yukawa couplings among chiral fermions localized at the intersections of two (stacks of) branes can only be generated by a physical effect that is non-local, and it is therefore suppressed by the exponential of the area of a string worldsheet stretched between the intersection points \cite{Cvetic:2003ch, Abel:2003vv}, $\exp(- L^2 / l_s^2)$.
These two examples qualitatively agree with the general expectation discussed in the previous paragraph. In the former, the Weak Gravity Conjecture (to be discussed in the next subsection) implies a general lower bound on a gauge coupling of the form $g \gtrsim \Lambda / M_\text{Pl}$, where in the presence of axion couplings $\Lambda$ can be interpreted as a quantum gravity cutoff scale~\cite{Heidenreich:2021yda}. In the latter, the overall volume of the compact dimensions must satisfy $V \gtrsim L^2 l_s^4$, and therefore $\exp(- L^2 / l_s^2 ) \gtrsim \exp( - \mathcal{O}(1) \times M_\mathrm{Pl}^2/\Lambda_\text{QG}^2 )$ by virtue of the relationship $M_\text{Pl}^2 \sim l_s^{-8} V$ and after identifying $\Lambda_\text{QG} \sim l_s^{-1}$.

The examples of the previous paragraph highlight two expected properties of global symmetries in a theory of quantum gravity: (1) that not all approximate symmetries are created equal; and (2) that approximate symmetries come at a cost. We turn to these topics next.

\subsection{The cost of approximate global symmetries}

The discussion of the previous section suggests that a general consequence of approximate global symmetries is a lowering of the cutoff scale as $\sim M_\text{Pl} / \sqrt{\log x^{-1}}$, with $x$ some parameter such that the limit $x \rightarrow 0$ restores the corresponding symmetry. For example, $x = y_f$ for a fermion chiral symmetry, or $x = \mu^4 / M_\text{Pl}^4$ for the shift-symmetry of an axion, with $\mu^4$ the contribution to the axion potential.  This conclusion seems a reasonable lower bound on the `cost' of realizing an approximate global symmetry, but it is not the whole story.

An alternative realization of an approximate global symmetry can be obtained by taking the weak coupling limit of a gauge theory. In this case, the obstruction to recovering an unbroken global symmetry is embodied by the Weak Gravity Conjecture (WGC) \cite{ArkaniHamed:2006dz} (see e.g.~\cite{Harlow:2022gzl} for a recent review). Loosely speaking, the WGC implies the presence of `super-extremal' degrees of freedom, whose charge-to-mass ratio is larger than that corresponding to an extremal black hole. Perhaps the most notable consequence of this statement is the existence of an upper bound on the cutoff scale of a $U(1)$ gauge theory given by $\Lambda \lesssim g M_\text{Pl}$, related to the presence of new dynamics linked to the existence of magnetic monopoles that satisfy the super-extremality bound \cite{ArkaniHamed:2006dz}. A more sophisticated version of this statement, but that nevertheless seems to be satisfied in all known examples, suggests that a full \emph{tower} of super-extremal states should be part of the theory \cite{Heidenreich:2015nta,Heidenreich:2016aqi,Montero:2016tif,Andriolo:2018lvp}. In combination with the species bound (the expectation that gravity becomes strong at the scale $\Lambda_\mathrm{QG} \sim M_\mathrm{Pl}/\sqrt{N}$~\cite{Veneziano:2001ah, Dvali:2007hz} in a theory with $N$ degrees of freedom), this translates into $\Lambda_\text{QG} \sim g^{1/3} M_\text{Pl}$~\cite{Heidenreich:2016aqi,Heidenreich:2017sim}. (Remarkably, the same $g^{1/3}$ scaling on the cutoff appears from an entirely different argument involving massive Stueckelberg gauge fields~\cite{Craig:2019zkf}.) In other words, the cost of approaching a global symmetry by taking the $g \rightarrow 0$ limit of a gauge theory appears to be power-law rather than logarithmic. This is a much more severe cost, and a stronger constraint on model-building.

It is important to assess whether it is really correct that the cost of a small gauge coupling in quantum gravity is always power-law, rather than logarithmic. A possible loophole arises from the `clockwork' mechanism, in which an exponentially large integer arises as a product of smaller integers~\cite{Choi:2014rja, Choi:2015fiu, Kaplan:2015fuy}. In the present context, clockwork means that if one higgses a set of $N$ gauge fields down to the diagonal, one can obtain a small IR gauge coupling of order $q^{-N}$ times the UV gauge coupling ($q$ a small integer), possibly evading the WGC constraint even if the UV theory obeys the WGC~\cite{Saraswat:2016eaz}. Whether such a mechanism  can be realized in actual UV-complete quantum gravity theory is unclear, and deserves attention.

All in all, we can summarize many of these developments by saying that quantum gravity often does not {\em forbid} ingredients one might want for model-building, but it makes them come at a cost. Often, this cost is that the theory breaks down at an energy $\Lambda_\mathrm{QG} \ll M_\mathrm{Pl}$. How fast that occurs will depend on the specific manner in which we are approaching the limit of an unbroken global symmetry.

\subsection{Holographic entropy}

One of the properties of gravitational theories that most sharply brings into question the validity of EFT even at low energies is the observation that the entropy of gravitational systems appears to be sub-extensive, scaling according to an area law. This applies even to black holes in otherwise asymptotically flat space, whose entropy is given by the Bekenstein-Hawking formula \cite{Hawking:1975vcx,Bekenstein:1973ur}
\begin{equation}
	S_\text{BH} = \frac{A}{4 G_N} \sim L^2 M_\text{Pl}^2 .
\end{equation}
Indeed, it is expected that this is an upper bound on the entropy of any system that gravitates: trying to exceed this bound would result in the system collapsing into a black hole \cite{Bekenstein:1980jp,Bousso:1999xy}.

This is obviously at odds with our understanding of entropy in the context of EFT, where entropy is an extensive quantity proportional to the volume of the system, and that may be as large as $S \sim L^3 \Lambda^3$, with $\Lambda$ a UV-cutoff (e.g.~for a gas in a box we might estimate $S \sim L^3 T^3$, and we might take the temperature to be as large as the cutoff up to which we can treat the system as a gas of non-interacting point particles).

It is therefore clear that for a fixed UV-cutoff the validity of the EFT description must break down for sufficiently large $L$ -- that is, far enough into the IR. This is one of the most concrete examples of how the idea of decoupling between UV and IR that is central to the EFT framework breaks down in the presence of gravity. A more concrete, semi-quantitative lesson that stems from this observation is that local quantum field theory over-counts the number of degrees of freedom in gravitational systems. In the most extreme case where the system under consideration is our own observable Universe, this observation can have profound implications for our understanding of the cosmological constant problem -- both how we formulate it and what it would mean to solve it. This will be discussed in section IV.2.

\subsection{The Swampland program}

Some of the ideas discussed in the previous subsections, like the absence of global symmetries of the holographic entropy bounds, go back to the 1970s. In recent times, it is common to loosely refer to these type of considerations as being part of the so-called Swampland Program, after the term introduced in \cite{Vafa:2005ui}. This has become an umbrella term to refer to the effort of trying to understand how gravitational considerations can affect physics at low energies, with the goal of discriminating between those EFTs that belong in the \emph{landscape} of theories that can descend from a theory of quantum gravity, and those in the \emph{swampland} of theories that do not. (For recent reviews, see~\cite{Brennan:2017rbf,Palti:2019pca,vanBeest:2021lhn,Grana:2021zvf}.)

A promising effort in this direction has been to investigate patterns in specific string constructions with the hope of discovering some underlying principle behind these constraints. A concrete example of this takes us back to the Tower WGC discussed in III.2. Indeed, the expectation that an infinite tower of states becomes light in the limit that a global symmetry is restored, as suggested by the conjecture, is not expected to be an isolated incident, but rather representative of a more general property of theories of quantum gravity, namely that limits in which an approximate global symmetry becomes exact are screened. In particular, one finds that they lie at {\em infinite distance} in field space. The so-called Swampland Distance Conjecture (SDC) makes this observation precise, stating that whenever a scalar field $\phi$ traverses a large (geodesic) distance $\Delta \phi$ homogeneously in field space, an infinite tower of modes become  exponentially light, with $m \propto \exp(- O(1)  \times \Delta \phi/M_\mathrm{Pl})$~\cite{Ooguri:2006in} (see also~\cite{Klaewer:2016kiy}). Kaluza-Klein modes are a familiar example, with a tower of KK modes of the graviton having mass $\propto 1/R$ and the canonically normalized radion field $\phi \sim \log|R|$. There are also limitations on localized large field excursions which tend to manifest in gravitational collapse or other instability rather than the appearance of a mode tower~\cite{Nicolis:2008wh,Draper:2019utz,Draper:2019zbb,Dolan:2017vmn}.  These breakdowns of large field distance limits in quantum gravity may have interesting phenomenological applications, 
but we would emphasize that even ``large-field inflation,'' where observable tensor modes  could be correlated with a Planck-scale field space distance, only requires (to fit cosmological data) that $\Delta \phi \sim O(1) \times M_\mathrm{Pl}$ and could be entirely consistent with this bound. 

Another phenomenologically relevant expectation about quantum gravity is the Completeness Hypothesis, which asserts that a consistent theory must contain objects of every charge allowed by gauge invariance~\cite{polchinski:2003bq}. One immediate consequence is that magnetic monopoles should exist. It is now understood that the Completeness Hypothesis follows from the absence of generalized (non-invertible) global symmetries~\cite{Gaiotto:2014kfa, Harlow:2018tng,Rudelius:2020orz, Heidenreich:2021tna}. In theories where the gauge group has discrete structure, completeness appears linked to the WGC~\cite{Craig:2018yvw}, and can have interesting implications such as the existence of novel types of cosmic strings.

Although exploring the growing diversity of string theory ``data'' in search for underlying patterns may be a hopeful endeavor, it is hard to ignore that the earliest insight for the best-motivated conjectures, such as the absence of global symmetries or the holographic entropy bounds, actually originated from studying the properties of black holes. Exploring the properties of these fascinating objects under a diversity of circumstances in search for additional intuition has seen renewed interest in recent times; see, e.g., the recent review~\cite{Almheiri:2020cfm} and the Snowmass white papers~\cite{Bousso:2022ntt,Blake:2022uyo,Bena:2022ldq}.
Perhaps not coincidentally, recent progress on the black hole information puzzle forgoes input from perturbative string theory, relying instead on semiclassical gravity techniques \cite{Penington:2019npb,Almheiri:2019psf}, and renewing hopes that constraints such as the absence of global symmetries could be understood using similar techniques \cite{Harlow:2020bee,Chen:2020ojn,Hsin:2020mfa}. It is hard to imagine that further exploring the properties of black holes would not yield even more surprising outcomes.

\section{Particle physics beyond EFT}
\label{sec:particlephysics}

\subsection{The strong CP problem and axions}

The absence of global symmetries in a theory of quantum gravity forces a reevaluation of certain extensions of the Standard Model featuring very good approximate global symmetries. A well-known example is the Peccei-Quinn solution to the strong CP problem, which requires a global $U(1)_\text{PQ}$ symmetry to remain unbroken to an exceptional degree. In particular, contributions to the axion potential other than those from QCD must remain subdominant by at least 10 orders of magnitude. This lack of robustness against additional contributions to the axion potential goes by the name of the ``axion quality problem''. Gravitational effects are expected to provide one such irreducible source of PQ symmetry-breaking.

One can take two complementary views in light of this tension. First, it is an old observation that axions are a common occurrence in string theory constructions \cite{Witten:1984dg}. Upon compactification, the higher-form antisymmetric tensor fields present in any string theory lead to a large number of pseudoscalar zero-modes, as determined by the topology of the compact manifold. Moreover, recent arguments also suggest that the existence of axions may actually be necessary from general principles in quantum gravity, and not a lamppost effect, as they sometimes play a role in eliminating would-be global symmetries~\cite{McNamara:2019rup,Heidenreich:2020pkc}. It is natural to wonder whether one of these axions could be \emph{the} QCD axion. For this to be the case, the axion would need to avoid becoming heavy already at tree-level due to other ingredients also generic in higher dimensional constructions such as fluxes and branes \cite{Douglas:2006es,McAllister:2008hb}. Similarly, \emph{all} non-perturbative contributions to the axion potential would need to be extremely tiny -- not just those from other gauge instantons, but also string instantons \cite{Dine:1986zy}, gravitational instantons \cite{Coleman:1989zu,Kallosh:1995hi}, or D-branes wrapping the corresponding cycle \cite{Becker:1995kb}. On the other hand, all of these effects are exponentially small, so string theory has in some sense taken the logarithm of the axion quality problem~\cite{Svrcek:2006yi}. Furthermore, these instanton effects are not all independent; for instance, the Yang-Mills instantons localized on a stack of D$p$-branes are continuously connected to Euclidean D$(p-4)$-brane instantons wrapping the same cycle~\cite{Witten:1995gx, Douglas:1995bn}, which therefore should not be expected to introduce new dominant contributions that spoil the Strong CP solution. Studies of concrete examples indeed find that string theory can produce high-quality QCD axions that solve the Strong CP problem~\cite{Demirtas:2021gsq}. Such a stringy QCD axion is expected to be situated in a sector with many axion-like particles~\cite{Arvanitaki:2009fg}, with highly structured couplings that conspire to eliminate many potential global symmetries~\cite{Heidenreich:2020pkc}.

This observation links the resolution of the strong CP problem to the existence of a plenitude of axions -- with the cosmological and observational consequences that entails. From CMB polarization measurements to the properties of the matter-power spectrum to gravitational waves signal, this is an example of how considerations beyond the realm of EFT can motivate vibrant  phenomenology.

An alternative approach is to focus on solutions to strong CP that do not require the appearance of high quality global symmetries. In light of current constraints, extensions of the Standard Model that restore parity symmetry \cite{Babu:1988mw,Babu:1989rb,Barr:1991qx} have been argued to be highly promising \cite{Craig:2020bnv}. As recently discussed in~\cite{Craig:2020bnv}, successful parity solutions rely on a see-saw implementation of the masses of all the light SM fermions. An irreducible signature is the presence of $W'$ and $Z'$ gauge bosons, with collider searches providing the most stringent test, and setting a lower bound on the parity-breaking scale $v' \gtrsim 18$ TeV~\cite{ATLAS:2019erb,ATLAS:2019lsy}. A future high energy collider will probe parity-breaking scales well above  $100$ TeV, and would be a decisive test of these solutions.
Beyond colliders, the expected breaking of parity due to gravitational interactions highlights additional possibilities to probe these models. For example, parity-breaking dimension-5  Planck-suppressed operators generate a non-zero $\theta$ that is small enough to satisfy current constraints, but large enough to be measurable in short- and mid-term future experiments if the breaking is maximal. If other sources of parity violation are present, they can disentangled from $\theta$ by correlated EDM measurements, providing experimental evidence against parity solutions~\cite{deVries:2018mgf,deVries:2021sxz}. Complementarily, if the explicit breaking was small, the production and collapse of domain walls in the early universe (associated with the spontaneous and explicit breaking of parity respectively) leads to a stochastic background of gravitational radiation potentially observable in current and future low-frequency gravitational wave experiments. A discovery on either of these fronts would provide circumstantial evidence in favor of parity-solutions complementing colliders searches~\cite{Craig:2020bnv}. 

\subsection{Naturalness and the weak scale}

Some of most puzzling problems of the Standard Model are problems of naturalness -- chief among them the cosmological constant and electroweak hierarchy. The framework of effective field theory is at the heart of the way that we set to formulate and address these puzzles. In particular, the decoupling between UV and IR scales that is built into the EFT toolkit makes us expect that features of at theory that are accessible from the IR should find their origin at commensurately low energies. This logic leads to the conclusion that a mechanism of spontaneous symmetry breaking that explains the observed value of the weak scale must make itself manifest at scales $\Lambda \sim 4 \pi v \lesssim$~TeV. The vast majority of attempts to address the electroweak hierarchy problem over the last few decades have pursued this logic.

Gravitational considerations offer an alternative approach to this problem, by virtue of the  UV/IR mixing of scales characteristic of gravity. The challenge is of course to implement this idea in a way that concretely addresses the electroweak hierarchy, and that opens a window for experiments to probe or at least constraint this possibility. Attempts along these lines include leveraging the WGC to set an upper bound on the weak scale of any consistent theory in which the mass of WGC-satisfying states is related to the Higgs vev~\cite{Cheung:2014vva,Craig:2019fdy}, a possibility that can lead to modifications in Higgs properties and interesting dark matter dynamics~\cite{Craig:2019fdy}. Other suggestions include leveraging the presumed absence of stable non-supersymmetric AdS vacua~\cite{Ooguri:2016pdq}, together with the existence of such vacua in compactifications of the Standard Model or its extensions~\cite{Arkani-Hamed:2007ryu}, to constrain neutrino masses~\cite{Ooguri:2016pdq,Ibanez:2017kvh} and, in turn, the Higgs vev \cite{Ibanez:2017oqr,Gonzalo:2018tpb}. A difficulty with such arguments is that there no way to demonstrate that the AdS compactifications in question are actually stable, so there is no sharp conflict with the conjecture.

At the moment, none of these examples provide what can be considered a bona-fide solution to the hierarchy problem, but they open a novel window for exploration.

\subsection{Dark photons}

In effective field theory, an abelian spin-1 boson can simply be given a mass, with no obvious pathology in the theory. This raises the question of whether the photon might have a mass. Experimental bounds on the photon mass are very strong, but within the framework of QFT, it is difficult to argue that the photon is truly massless. On the other hand, in string theory, vector boson masses not arising from the Higgs mechanism always appear in conjunction with more dynamics. We could consider writing the photon mass in the form of a BF theory, i.e., with a coupling $B \wedge F$ between a dynamical 2-form gauge field and the photon field strength. In QFT, such a Stueckelberg theory is entirely equivalent to simply writing a mass term. However, if we take the existence of the 2-form field $B$ literally, in the gravitational context we can apply the WGC to  infer the existence  of  a string charged under $B$, with tension $T \lesssim f M_\mathrm{Pl}$ where  the photon mass is $e f/(2\pi)$. If this theory does not simply take the form  of  a Higgs mechanism, then the charged strings are fundamental  objects (not solitonic solutions of the low-energy EFT), and local QFT breaks down at the scale  of the string tension: 
\begin{equation}
M_\mathrm{string} \lesssim 2\pi \sqrt{\frac{m_\gamma M_\mathrm{Pl}}{e}}.
\end{equation}
This translates the  experimental bound on the  photon mass into an upper bound on the energy  at  which QFT is valid. This bound turns out to be  well {\em below}  the LHC energy where QFT has been  observed to work.  Thus, we conclude that the photon is truly massless~\cite{Reece:2018zvv}.   Variants of this argument can also significantly constrain theories of dark photon dark  matter.

This argument has a crucial loophole: if the true quantum of electric charge is actually many orders of  magnitude below the electron charge, then the coupling $e$ above  is replaced by $e/N$, which can change the conclusion. This may seem unnatural,  but such a  large integer could arise  in a  concrete  model as a product of  smaller integers~\cite{Craig:2018yld}. This is yet another application of clockwork~\cite{Choi:2014rja, Choi:2015fiu, Kaplan:2015fuy}, and  again points  to the importance  of understanding possible charge assignments arising for  light fields in quantum gravity.

\subsection{Nature of weak-coupling limits}

The SDC and Tower WGC state the weak-coupling limits in quantum gravity always involve a tower of modes becoming  light. In known theories of quantum gravity, these  towers  take on very specific forms. Either extra dimensions decompactify, so that the tower consists  of Kaluza-Klein modes, or a fundamental string exists with a tension that becomes small  in the weak-coupling  limit, so that the tower consists of string oscillator modes. The Emergent String Conjecture states that these are the {\em only} possibilities for the behavior of the lightest tower~\cite{Lee:2019wij}, based on geometric arguments in F-theory  compactifications~\cite{Lee:2018urn, Lee:2019xtm}. It has been  further suggested that, at  least in 4d ${\cal N}=1$ supersymmetric compactifications, there is always a low-tension, axionic string in every  weak coupling limit~\cite{Lanza:2020qmt, Lanza:2021udy}; in  the decompactification limit, the string modes are heavier than the KK modes, but  they still exist. Furthermore, the existence of an axion coupling in general indicates the tower should be stringy~\cite{Heidenreich:2021yda}.

The two possible weak-coupling scenarios have very different implications for the quantum gravity cutoff energy $\Lambda_\mathrm{QG}$. A tower  of charged modes at the scale $g M_\mathrm{Pl}$ does not necessarily invalidate local QFT at that scale: in the Kaluza-Klein case, the true quantum gravity cutoff is the higher-dimensional Planck scale, as large as $g^{1/3} M_\mathrm{Pl}$, not the KK  scale.  However, a tower of fundamental string modes does  indicate the complete breakdown of local QFT at $g M_\mathrm{Pl}$. 

If true, the Emergent String Conjecture has powerful phenomenological consequences. Let us give a hypothetical example, for illustration. Suppose that a future precision experiment discovered a very weakly-coupled gauge field, e.g., an $L_\mu - L_\tau$ gauge boson with coupling $g_L  \sim 10^{-6}$. Because there is chiral matter charged under this gauge field, it could not arise as a Kaluza-Klein $U(1)$ from a circle compactification. The Emergent String Conjecture would then suggest that there are fundamental strings at or below the scale $g_L M_\mathrm{Pl} \sim 10^{12}\,\mathrm{GeV}$. This would be a major restriction on theories of UV physics, ruling out conventional GUTs, models of high-scale inflation, and some models of neutrino masses  and leptogenesis, for example. Conversely, a discovery of such a weakly coupled gauge boson together with evidence that local field theory functions at high scales (such as a measurement of inflationary tensor modes) could empirically falsify the Emergent  String Conjecture in our universe.

The Emergent String Conjecture  is  one of the most promising ideas for connecting generalities about quantum gravity with concrete experimental developments. Obtaining a convincing argument for the conjecture from general principles, rather than the F-theory lamppost, is a  high priority. Alternatively, one could search for counterexamples of consistent quantum gravity theories that violate this conjecture. Such investigations are only beginning,  but have great promise.

\section{Quantum gravity and dark  energy}
\label{sec:darkenergy}

Our universe is undergoing accelerated expansion. The simplest explanation is that the universe is tending toward a late-time de Sitter phase, with a positive cosmological constant of magnitude $\Lambda_\text{CC} \sim(10^{-33}\,\rm{eV})^2$. 
We are compelled to address the questions of why it is positive and why it is so small compared to other scales in fundamental physics. The landscape paradigm has long been suggested as a possible picture for the second question, but recently the swampland program has pushed back in interesting ways and stimulated new ways of thinking about both problems. 

\subsection{Quantum gravity in de Sitter space}

A consistent theory of quantum gravity with a de Sitter-like solution is probably essential to fully answer these questions. However, compared to asymptotically flat or anti de Sitter space, where supersymmetric string theory solutions and the AdS/CFT correspondence provide well-understood, quantitatively predictive models, quantum gravity in de Sitter space remains relatively less understood. 
One line of research with a long history is to consider the effective field theory of moduli in different string compactifications and construct the effective potential generated by fluxes, branes, quantum corrections, and other effects, searching for minima of positive energy density. 
This String Landscape potential could admit a gigantic number of stationary points, and perhaps the probability of finding a local minimum with the correct magnitude can be O(1). The reason we live in it could be a selection effect. The belief that at least some elements of this picture are relevant to explain the cosmological constant problem is widely held~\cite{Bousso:2000xa}.

Constructing de Sitter vacua in string models in a completely controlled way has proven to be a subtle task. From the outset, one must confront the problem that in the asymptotics of moduli space, which are otherwise under the best theoretical control, the potentials are generally falling rapidly and monotonically to zero~\cite{Dine:1985he}. Furthermore, various no-go theorems cut away otherwise natural settings~\cite{Maldacena:2000mw,Hertzberg:2007wc}. However, these observations do not cover all the scenarios and ingredients at the string theorist's disposal, and exciting progress continues to be made, both in developing new tools and constructions, and in putting older constructions under the microscope. The literature is as substantial as it is rich and we will not attempt to do justice to it here, but for two recent examples, we note the development in~\cite{Demirtas:2019sip} of new techniques to obtain the small superpotential used in the KKLT construction~\cite{Kachru:2003aw}, and of new constructions using hyperbolic compactifications  in M-theory~\cite{DeLuca:2021pej}.


It is therefore quite reasonable to think that among these constructions, however constrained they might be, there could be the right ingredients to describe our universe, and in addition some realization of the landscape paradigm to address the cosmological constant problem. However, as this whitepaper is about the swampland program, we will focus on two alternative (and more qualitative) ideas which have been proposed. The first is simply that theories with effective potentials that exhibit de Sitter minima are part of the swampland~\cite{Danielsson:2018ztv,Obied:2018sgi}. The other is that we cannot decide the status of de Sitter vacua in quantum gravity one way or the other from perturbative string constructions, even if such effective potentials can be constructed.  

The first idea was formalized in what are known as de Sitter conjectures~\cite{Obied:2018sgi,Garg:2018reu,Ooguri:2018wrx}, and they have stimulated much activity in recent years. These conjectures bound the first and second derivatives of the scalar potential. The refined de Sitter conjecture of~\cite{Ooguri:2018wrx} states that, in Planck units and at every point in  the domain, either the slope of the potential is greater than the potential itself, or it has a negative curvature in some direction which is greater in magnitude than the potential. The inequalities are conjectured to hold up to some unknown coefficients. They are inconsistent with metastable de Sitter minima and very flat regions of scalar potentials, so they imply both interesting constraints on the inflationary potential, and that different ingredients are needed to explain dark energy. There could, for example, be weak time dependence from rolling in a shallow scalar potential, as in models of quintessence. Such scenarios can be probed by cosmological measurements of the time dependence of the equation of state parameter, from which bounds on the unknown coefficients in the inequalities can be obtained, and the tensor-to-scalar ratio gives another observable handle~\cite{Obied:2018sgi}. Future cosmological measurements will have the sensitivity to place strong bounds on the coefficients in the inequalities (see, e.g.~\cite{Dias:2018ngv,Heisenberg:2018yae,Kinney:2018nny,Akrami:2018ylq,Agrawal:2018rcg,Raveri:2018ddi}). In the last few years a large body of literature has explored the implications for models of inflation and quintessence. Furthermore, variations on the de Sitter conjectures have suggested that even if metastable dS vacua can exist, they must be very short-lived~\cite{Bedroya:2019snp}. Both spontaneous decompactification~\cite{Giddings:2003zw,Giddings:2004vr} and spontaneous compactification~\cite{Dine:2004uw,Draper:2021qtc,Draper:2021ujg} processes could lead to lifetimes not much longer than a few Hubble times for suitable parameters.

The second idea was reviewed and expanded in~\cite{eftmyths}. One line of argument from~\cite{eftmyths} goes as follows. If there was a dS minimum of a potential arising from a string construction, it would be local rather than global. Consider the case where there is another isolated minimum which is stable due to supersymmetry (similar conclusions hold for stable vacua at infinity.)  On one hand, the CDL decay of the metastable dS vacuum does not produce the flat solution corresponding to the stable minimum, but rather a cosmology with curvature, and it is not known how to formulate quantum gravity on the CDL universe (the S-matrix is unavailable~\cite{Bousso:2005yd}. On the other hand, an experimenter in asymptotically flat space is also unable set up a large (radius $>$ 1/Hubble) region locally approximating the metastable dS vacuum: such a region collapses to a black hole; observers inside reach the singularity in O(1) Hubble times; and the decay of the black hole is unrelated (and much faster) than the CDL decay of the false vacuum. This argument and various others of the same type support the conclusion that different solutions of effective field theories, even static ones, need not all correspond to states in the same (or any) model of quantum gravity. In perturbative string theory constructions the EFT is formally derived by matching onto flat space scattering amplitudes or AdS boundary correlators, and so it is possible that Minkowski or AdS solutions of the EFT make perfect sense, while no conclusion can be drawn either way about the status of dS minima or other solutions with different asymptotics.

Although these two ideas are incompatible with each other, they articulate important and, in a sense, radical ideas that should drive theoretical progress. The de Sitter conjectures, if true, greatly impact cosmological model building and the possible futures for the universe.  If false, they represent quantitative targets for string constructions beyond just the existence of a vacuum with a small positive energy density. If the second idea is correct, it suggests that rather different ingredients may provide more natural descriptions of de Sitter quantum gravity. With this in mind let us briefly note a number of interesting directions that have seen recent process.

One line of semiclassical reasoning suggests that de Sitter holography should be formulated as a matrix quantum mechanics, with a large but finite-dimensional Hilbert space~\cite{Banks:2006rx,Banks:2020zcr,Susskind:2021dfc}.
This might account for the curious property that local excitations in de Sitter space lower the total entropy~\cite{Banks:2006rx} 
(see also~\cite{Dinsmore:2019elr,Susskind:2021dfc}), which was given a path integral description in terms of constrained states in~\cite{Draper:2022xzl}. Recent work~\cite{Anninos:2020geh, Anninos:2021eit} discusses possible matrix model descriptions of dS${}_2$. Along complementary lines there has also been rapid progress in the construction of holographic dualities for dS${}_3$, utilizing generalizations of the $T\bar T$ deformation~\cite{Gorbenko:2018oov,Shyam:2021ciy,Coleman:2021nor} (see also~\cite{Anninos:2021ihe}). Other interesting work~\cite{Susskind:2021dfc,Shaghoulian:2021cef,Shaghoulian:2022fop} has proposed minimal surface-type prescriptions~\cite{Ryu:2006bv} for geometrically computing entanglement entropies in de Sitter and its excitations, and the thermodynamics of subregions in dS${}_4$ was recently analyzed semiclassically in~\cite{Draper:2022ofa,Banks:2020tox}. These are exciting developments that could lead to a complete understanding of quantum gravity models relevant for the universe we live in. For particle physics, progress in understanding how holography operates in the static patch of de Sitter space and in other causal diamonds could lead to new ways of reconciling effective field theory with a small cosmological constant, to which we now turn.

\subsection{The cosmological constant problem and entropy bounds}

 It is an old idea that entropy bounds~\cite{Fischler:1998st,Bousso:1999xy,Bousso:1999cb} can be used to neutralize the standard cosmological constant fine-tuning problem suggested by effective field theory reasoning~\cite{ckn,Thomas:2002pq}. While not a completely conventional ingredient in the swampland toolkit, there is a growing sense that entropy considerations will play an important role in the program in the future (in fact the refined dS conjecture was based on entropic arguments~\cite{Ooguri:2018wrx}), and moreover they fit naturally under same umbrella in which UV physics affects IR physics in un-EFT-like ways (known broadly as ``UV/IR mixing.'') Recently there has been a burst of new activity related to the Cohen-Kaplan-Nelson (CKN) bound~\cite{ckn,Bramante:2019uub,Banks:2019arz,Cohen:2021zzr,Davoudiasl:2021aih,Blinov:2021fzl,Ramakrishna:2021sll}. 

The CKN bound can be formulated as the observation that a high-entropy state of a field theory tends to undergo gravitational collapse when its entropy is only of order $A^{3/4}$ in four dimensions, rather than $A$, assuming that the number of fields is fixed~\cite{ckn,eftmyths,Banks:2019arz}.  Formulated in terms of UV and IR cutoffs, the CKN bound takes the form $\Lambda_{IR} > M_p/\Lambda_{UV}^2$. Resolving the appropriate interpretation of this UV/IR correlation is an ongoing effort of considerable importance for phenomenology; depending on the meaning, it can either imply that quantum gravity effects are detectable in precision experiments~\cite{Bramante:2019uub,Cohen:2021zzr,Davoudiasl:2021aih} or that they are as negligible as one might expect from an effective field theory supplemented by corrections in powers of $G_N$~\cite{Banks:2019arz,Blinov:2021fzl}.

However, even the more ``pessimistic'' interpretation of~\cite{Banks:2019arz,Blinov:2021fzl} still allows relevance for the cosmological constant fine-tuning problem. The idea is that not all of the high-energy degrees of freedom in  an EFT represent truly independent degrees of freedom of the quantum gravity model of which the EFT is a low-energy approximation. This is already suggested by the area law, but the CKN bound is stronger, applying specifically to degrees of freedom which can be well-approximated by local field theory. Since the quantum contribution of high energy field-theoretic degrees of freedom to the effective cc is the source of its apparent fine tuning, depleting them in a suitable way could resolve the problem. Lacking a theory of quantum gravity in de Sitter space, it is not known how this depletion might manifest, and in general it is a subtle problem how holography yields bulk theories with a lot of spacetime symmetry. However~\cite{Blinov:2021fzl} developed a rough method for estimating the effects of a scale-dependent depletion of the degrees of freedom on the cc and other observables using finite-volume techniques, and the results of this na\"ive approach suggest that (1) in principle the fine-tuning problem can be addressed without having a large effect on most IR physics, and (2) the depletion needs to be of the somewhat stronger CKN form than that implied by the Bekenstein bound. 

Going forward, it is imperative to understand how a holographic theory of quantum gravity in de Sitter space would impact the cosmological constant problem, 
to understand better the meaning of CKN bound and similar thought experiments,
and to fully explore the range of applications of entropy bounds to the swampland program. These questions pose exciting opportunities for high energy theory at the interface of quantum gravity and phenomenology.

\section{Coda: The perils of technical naturalness}
\label{sec:conclusions}

Many of the outstanding puzzles of the Standard Model are ``hierarchy problems,'' in the general sense that they ask why a dimensionless number is small  rather than $O(1)$, which might be viewed as the generic expectation. These include the famous electroweak hierarchy problem ($m_h^2/M_\mathrm{Pl}^2 \ll 1$); the cosmological constant problem ($\rho_\Lambda/M_\mathrm{Pl}^4 \ll 1$); the strong CP problem (${\bar \theta} \ll 1$); and the flavor problem ($y_f \ll 1$, or more broadly, the peculiar structure of fermion masses and mixings). While all of these have been treated as serious puzzles and have received sustained attention from particle theorists over decades, the first two have generally been treated as more serious problems than the latter two  puzzles because they are problems of {\em technical naturalness}: that is, they are problems in which loop corrections to the small quantities in question exhibit power divergences, and so a theory that might appear to predict a small value at some loop order can be completely spoiled by higher-order corrections. From this viewpoint, a small Yukawa  coupling is not viewed as a technical hierarchy problem, because loop corrections are proportional to the Yukawa itself. One can simply assert that it is small, and further calculation will not undermine this assertion. Technical naturalness can be an important criterion to keep in mind when proposing a solution to a hierarchy problem, but its use in the community has become dogmatic, at times to the point of absurdity, eclipsing any other considerations about a theory's aesthetic appeal or conceptual soundness. Even as too much devotion to the criterion of technical naturalness has led to highly engineered and artificial proposals for the electroweak hierarchy and the cosmological constant problem, it has also tended to diminish the importance of the flavor puzzle and (perhaps to a lesser extent) the strong CP problem in the minds of the particle theory community.

Quantum gravity provides us with powerful reasons for seeking theories that are truly natural, in the sense that they explain small dimensionless numbers in terms of $O(1)$ inputs. As an example, consider the flavor symmetries of the Standard Model. All of the Standard Model Yukawa couplings, except the top Yukawa, are small numbers, so there is a large approximate flavor symmetry at low energies; for instance, there is an approximate $SU(3)_d$ symmetry rotating the three down-type quarks into each other. We expect that this symmetry should be badly broken at the quantum gravity cutoff energy. Traditional models that solve the flavor problem accomplish this. For example, in Froggatt-Nielsen models, there is a $U(1)$ horizontal symmetry under which the different down-type quarks carry different charge. This symmetry can be gauged, in quantum gravity (or a discrete group accomplishing similar results can be gauged). Alternatively, the quarks could live at different locations in extra dimensions. In either case, from the UV perspective, the three down quarks would be fundamentally different objects; their similarities at low energies  would be an emergent, infrared property.

In summary, the expectation that there are no global symmetries in quantum  gravity, and that at the UV cutoff scale there are not even {\em approximate} ones, provides a strong impetus, beyond simple aesthetic considerations, for the traditional model-building goal of explaining small numbers of the  theory in terms of reasonable  inputs. Technical naturalness can be a useful guiding principle for IR effective field theories,  but a complete theory must go beyond it. This is not to say that a number cannot be small simply because of an accidental  cancelation, although quantum gravity may also constrain how many such fine-tunings can occur~\cite{Heckman:2019bzm}. It does mean that hierarchies that are not technically unnatural, like the flavor hierarchies, are important puzzles that could shed light on UV physics.

\section*{Acknowledgements} 
PD acknowledges support from the US Department of Energy under grant number
DE-SC0015655 and from the DOE Office of High Energy Physics QuantISED program under
an award for the Fermilab Theory Consortium ``Intersections of QIS and Theoretical Particle
Physics.''
The research of IGG is funded by the Gordon and Betty Moore Foundation through Grant GBMF57392, and by the National Science Foundation through Grant No.~NSF PHY-1748958.
MR is supported by the DOE Grant DE-SC0013607, the NASA Grant 80NSSC20K0506, and the Alfred P.~Sloan Foundation Grant No.~G-2019-12504.

\bibliography{whitepaper_refs}
\bibliographystyle{utphys}

\end{document}